\documentclass[prb,aps,amssymb,twocolumn,groupedaddress,notitlepage,superscriptaddress,longbibliography]{revtex4}
\usepackage{amssymb}
\usepackage{amsmath}
\usepackage{graphicx}
\usepackage{bm}
\usepackage[usenames,dvipsnames]{xcolor}
\definecolor{goodgreen}{rgb}{0.1,0.5,0}
\definecolor{goodred}{rgb}{0.7,0,0}
\usepackage[colorlinks,urlcolor=goodgreen,citecolor=blue,linkcolor=goodred]{hyperref}

\begin{document}

\title{A quantum-network approach to spin interferometry driven by Abelian and non-Abelian fields}

\author{A. Hijano}
\email{ahijano001@ikasle.ehu.eus}
\affiliation{Department of Applied Physics II, University of the Basque Country UPV/EHU, Apartado 644, 48080 Bilbao, Spain}
\affiliation{Donostia International Physics Center (DIPC), 20018 Donostia--San Sebasti\'an, Spain}
\affiliation{Centro de F\'isica de Materiales (CFM-MPC) Centro Mixto CSIC-UPV/EHU, E-20018 Donostia-San Sebasti\'an,  Spain}

\author{T. L. van den Berg}
\email{tineke.vandenberg@dipc.org}
\affiliation{Donostia International Physics Center (DIPC), 20018 Donostia--San Sebasti\'an, Spain}

\author{D. Frustaglia}
\email{frustaglia@us.es}
\affiliation{Departamento de F\'isica Aplicada II, Universidad de Sevilla, E-41012 Sevilla, Spain}

\author{D. Bercioux}
\email{dario.bercioux@dipc.org}
\affiliation{Donostia International Physics Center (DIPC), 20018 Donostia--San Sebasti\'an, Spain}
\affiliation{IKERBASQUE, Basque Foundation for Science, Euskadi Plaza, 5, 48009 Bilbao, Spain}

\begin{abstract}
We present a theory of conducting quantum networks that accounts for Abelian and non-Abelian fields acting on spin carriers. We apply this approach to model the conductance of mesoscopic spin interferometers of different geometry (such as squares and rings), reproducing recent experimental findings in nanostructured InAsGa quantum wells subject to Rashba spin-orbit and Zeeman fields (as, e.g., the manipulation of Aharonov-Casher interference patterns by geometric means). Moreover, by introducing an additional field-texture engineering, we manage to single out a previously unnoticed spin-phase suppression mechanism. We notice that our approach can also be used for the study of complex networks and the spectral properties of closed systems.
\end{abstract}

\maketitle

\section{Introduction}
Within the development of mesoscopic physics, coherent spin transport in quantum electronics has attracted a great deal of attention along the past decades.~\cite{Manchon_2015} This interest runs from the study of fundamental spin-based phenomena, such as weak antilocalization~\cite{WAL} and geometric/topological spin phases,~\cite{Banerjee_1996,Frustaglia_2020} to proposals for spintronic applications, such as spin field-effect transistors,~\cite{datta-das-90,Koo_2009,datta-das-15} spin filtering~\cite{Aharony_2008}, spin qubits~\cite{Foldi_2005,Kregar_2016} and, more recently, spin-based platforms for topological quantum computing.~\cite{Fu-Kane-09} Additionally, the understanding of quantum phenomena associated to coherent transport can pave the way to enhance and improve the sensibility of nanometer-size devices.~\cite{Batelaan_2009}
A common ingredient here is the role played by Abelian and non-Abelian phases produced by the carriers' spin dynamics under the action of magnetic textures originating from either (i) purely magnetic sources, such as micromagnetic arrays leading to inhomogeneous Zeeman coupling,~\cite{Loss_1990,Frustaglia_2001,Scheid_2006,Scheid_2007} (ii) purely electric sources, leading to spin-orbit interaction such as Rashba or Dresselhaus coupling,~\cite{Ramaglia_2006,Lia_2021} or (iii) hybrid sources, combining magnetic and spin-orbit fields.~\cite{Nagasawa_2012,Nagasawa:2013} 

Most experimental implementations are performed by using materials with strong Rashba spin-orbit coupling (RSOC) such as InAlAs/InGaAs heterostructures, which allow for the electrical control of the RSOC strength via top gates. The recent transport experiments with mesoscopic interferometers performed by Nitta's group~\cite{Nagasawa_2012,Nagasawa:2013,Nagasawa:2018,Wang_2019} showed evidence of electronic wavefunction manipulation through both the charge and the spin degrees-of-freedom. The former responds to magnetic fluxes whereas the latter reacts to the RSOC (and to additional Zeeman fields). The orbital coupling to magnetic fields gives rise to the Aharonov-Bohm effect~\cite{Aharonov_1959} and the spin coupling to electric fields is responsible for the Aharonov-Casher effect~\cite{Casher_1984} | the electromagnetic dual of the previous one. Experiments reporting the observation of the Aharonov-Casher effect in mesoscopic systems were carried out in HgTe heterostructures,~\cite{Koening_2006} three-dimensional topological insulators~\cite{Qu_2011} and Josephson junction circuits~\cite{Pop_2012}, whereas the Aharonov-Bohm effect has been extensively observed in metallic rings~\cite{Umbach_1986}, $p$-type GaAs heterostructures~\cite{Grbic_2007} and carbon nanotubes,~\cite{Bachtold_1999} among others.

The modelling of spin-dependent transport in mesoscopic systems can demand significant numerical efforts. The most popular technique has been the recursive construction of Green's functions from tight-binding Hamiltonians,~\cite{Thouless_1981,datta_1997} a very reliable method which sometimes turns out expensive in computational terms. More recently, the development of the open-source \textsc{Kwant} code,~\cite{Groth_2014} based on a wave-function scattering approach, represented a major step towards computational efficiency and stability with excellent results. Alternatively, particular systems characterized by an underlying network structure can be simulated by introducing more specific techniques based on a quantum-graph approach.~\cite{kowal_1990,Kostrykin_1999,Kostrykin_2000,Vidal:2000,Gnutzmann:2006,Bercioux:2004,Bercioux:2005A,Bercioux:2005B,Ramaglia_2006}

In this article, we apply the quantum-network approach to model the transport properties of mesoscopic spin interferometers subject to generic magnetic fields and RSOC by considering different geometries. The geometry of the interferometer is a key element, since the carriers' spin dynamics and the corresponding spin-phase gathering are sensitive to it.~\cite{Bercioux:2004,Bercioux:2005A,Bercioux:2005B,Ramaglia_2006,Koga_2006} Additionally, we propose a non-trivial extension of the quantum-network approach by considering the presence of in-plane Zeeman fields. We treat those in a similar way as the RSOC, however, they break time-reversal symmetry and, as a consequence the wave function describing each edge of the quantum-network cannot be written in a compact form as in the case of flux of a magnetic field~\cite{Kottos:1999} or a  RSOC.~\cite{Bercioux:2004}
The work is motivated by the experiments performed by Nitta's group \cite{Nagasawa_2012,Nagasawa:2013,Nagasawa:2018,Wang_2019} showing evidence of electronic spin manipulation by geometric means in Aharonov-Casher interferometry. These experiments were done by using two-dimensional arrays of $40\times40$ ring and square interferometers, facilitating the measurement of a self-averaged conductance: the presence of multiple but small imperfections in the individual ring and square geometries lead to a universal signal that is independent of those details. One experiment revealed the manipulation of the geometric spin phase independently of the dynamical spin phase in Rashba rings.~\cite{Nagasawa:2013} More recently, topological spin-phase transitions in polygonal Rashba circuits have been reported.~\cite{Wang_2019} By employing a quantum-network model, we reproduce the main results found in those experimental settings. Our approach provides a full quantum mechanical solution for the propagation of spin carriers inside the polygonal structure. The main strengths of our approach are the following: 
(i) we accounted for all possible propagating paths; 
(ii) there are no particular constraints imposed on the scattering matrix at the injector and collector nodes.

Moreover, we take some steps forward and identify novel interferometric characteristics by proposing a field-texture engineering. This results in a physical situation similar to the spin-helix effect arising in systems subjected to Rashba and Dresselhaus spin-orbit couplings (SOCs).~\cite{Schliemann:2003A,Schliemann:2003B,Koralek_2009} We also cross-check the results of our quantum-network approach by performing corresponding numerical simulations on tight-binding models.

The paper is organized as follows. In Sec.~\ref{formalisms} we take a quantum-network approach to model a one-dimensional  circular loop as a regular polygon with a large number of vertices by following the method described in Ref.~[\onlinecite{Bercioux:2005B}]. We generalize the method by introducing additional in-plane Zeeman fields beyond the perturbative approximation.~\cite{Nagasawa:2013} 
In Sec.~\ref{results} we derive the conductance as a function of the RSOC and Zeeman-field strengths and compare the obtained results to the experimental data and perturbative methods. Furthermore, we discover the occurrence of the transition line for which the Zeeman and Rashba terms cancel each other out and we study the effects on transport properties. We devote Sec.~\ref{res_tb} to compare the results obtained by the quantum-network method with those obtained by using a numerical tight-binding approach in the cases of clean and disordered systems. We present a short summary in Sec.~\ref{conclusions} where we analyze the strengths of the quantum network approach. In App.~\ref{appendix_R}, we propose a technical derivation of one of the major analytical results of this work.

\section{Model and Formalism}\label{formalisms}

\begin{figure}[!t]
    \centering
    \includegraphics[width=\columnwidth]{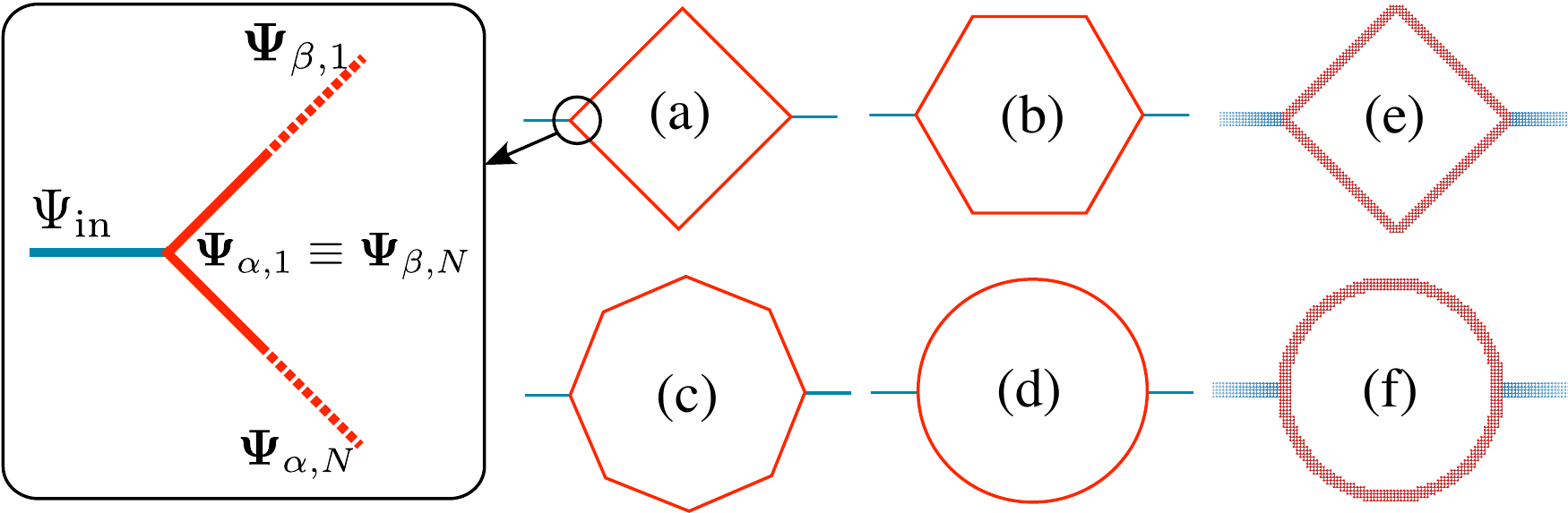}
    \caption{\label{polygons} Sketch of the polygonal structures that we consider for the quantum transport: (a) square, (b) hexagon, (c) octagon, (d) ring. In panels (e) and (f) we show the square and the ring of the discretized tight-binding version that we use in Sec.~\ref{res_tb}. The inset shows a zoom-in of the Y junction between the incoming lead and one of the vertices of the square.}
\end{figure}

We study the transport properties of ring-like structures by employing the formalism of quantum networks.  We focus on regular polygons and approximate ring geometries as polygons with a large number of edges --- see Fig.~\ref{polygons}(a-d). In general terms, a metric network (or graph), is a collection of nodes (or vertices), connected by edges (or one-dimensional intervals) of specified lengths.~\cite{Harrison_2010} In the graph terminology,  regular polygons are also known as \emph{2-regular graphs}.~\cite{Gnutzmann:2006} A quantum network is a metric graph equipped with a Schr\"{o}dinger operator.~\cite{Fulling:2007,Kottos:1999}
%
%
%
%
The wavefunction of the quantum network satisfies boundary conditions at the vertices, which ensure the continuity (uniqueness) of the wavefunction and the conservation of the probability current. The fulfilment of these boundary conditions guarantees that the resulting Schr\"{o}dinger operator is Hermitian.~\cite{Kostrykin_1999,Kostrykin_2000,Kottos:1999} We note in passing that this condition can be relaxed in order to account for non-Hermiatian Hamiltonians with $\mathcal{PT}$-symmetry.~\cite{Znojil_2009} The continuity condition implies that the wavefunction assumes a certain value at a vertex, regardless of the bond from which it is approached. The concept of extended normal derivative must be introduced to apply the second boundary condition, whose definition might vary depending on the differential operator, \emph{i.e.} it depends on the presence of a magnetic field or a SOC. In order to satisfy the current conservation condition, the sum of the outgoing extended derivatives at each vertex must vanish.~\cite{Kostrykin_1999,Gnutzmann:2006}
An extension for studying the $n$-particles quantum statistics has been proposed in Ref.~[\onlinecite{Harrison_2014}].

In the following subsections, we consider a quantum network composed of single mode quantum wires (QWs) subject to RSOC and to magnetic fields. In each case, we define a different Schr\"{o}dinger operator (Hamiltonian) and the corresponding extended derivative.
We start by introducing different spin-dependent Hamiltonians for the QWs composing the quantum network. Each Hamiltonian will be solved by using a spinorial plane wave ansatz of the following form
%
%
\begin{equation}\label{spinor}
\boldsymbol{\Psi}= \mathrm{e}^{\mathrm{i}kr}  \left(\begin{array}{c}
        \chi_\mathrm{A} \\ \chi_\mathrm{B}
    \end{array}\right),
\end{equation}
%
%
where $\chi_\mathrm{A/B}$ are the two components of the spinor, $k$ is the electronic momentum and $r$ is the local coordinate along the QW.

\subsection{Case of RSOC \& magnetic flux}

We start by considering a QW in the $xy$-plane  that points along the direction $\hat{\boldsymbol{\gamma}}=(\cos\gamma,\sin\gamma,0)$. The QW is subject to RSOC and a weak magnetic field perpendicular to the $xy$-plane that interacts with the spin carriers only through minimal coupling (\emph{i.e.}, no Zeeman coupling).~\cite{Bercioux:2004,Bercioux:2005A} The wire Hamiltonian reads
%
%
\begin{equation}\label{HamiltonianABC}
    \hat{\mathcal{H}}=\frac{1}{2m^*}(\boldsymbol{p}+e\boldsymbol{A})^{2}+\frac{\hbar k_{\mathrm{R}}}{m^*}\left[(\boldsymbol{p}+e\boldsymbol{A})\times\hat{\boldsymbol{z}}\right]\cdot\boldsymbol{\sigma}\; ,
\end{equation}
%
%
where $k_{\mathrm{R}}$ is the coupling constant of the RSOC (in inverse-length units), $\hat{\boldsymbol{z}}$ is the unit vector along the $z$-axis, $\boldsymbol{\sigma}$ is the vector of the Pauli matrices describing the electron spin, and $m^*$ is the electron effective mass. The RSOC strength $k_{\mathrm{R}}$ is related to the spin precession length  $L_{\mathrm{SO}}$ by $L_{\mathrm{SO}}=\pi/k_{\mathrm{R}}$. The wavefunction of the QW, fulfilling the Dirichlet boundary condition, can be written as a function of the values that the wavefunction takes at its vertices $\alpha$ and $\beta$:~\cite{Bercioux:2004,Bercioux:2005A}
%
%
\begin{align}\label{wavefunction1}
\boldsymbol{\Psi}(r)=&\frac{\mathrm{e}^{-\mathrm{i}\varphi(r)}\mathrm{e}^{-\mathrm{i}(\hat{\boldsymbol{\gamma}}\times\hat{\boldsymbol{z}})\cdot\boldsymbol{\sigma}k_{\mathrm{R}}r}}{\sin(k\ell)}[\sin{k(\ell-r)}\boldsymbol{\Psi}_\alpha %
\nonumber\\
&\hspace{1.5cm}+\sin{(kr)}\mathrm{e}^{\mathrm{i}\varphi(\ell)}\mathrm{e}^{\mathrm{i}(\hat{\boldsymbol{\gamma}}\times\hat{\boldsymbol{z}})\cdot\boldsymbol{\sigma}k_{\mathrm{R}}\ell}\boldsymbol{\Psi}_\beta],
\end{align}
%
%
where $\ell$ is the length of the QW,  $r$ is the local coordinate along the QW measured from vertex $\alpha$, and the momentum $k$ is related to the energy $\epsilon$ as $k=\sqrt{2m\epsilon/\hbar^2+k_{\mathrm{R}}^2}$. The spinors $\boldsymbol{\Psi}_\alpha$ and $\boldsymbol{\Psi}_\beta$ are the values of the wavefunction at the vertices $\alpha$ and $\beta$, respectively. 
In Eq.~\eqref{wavefunction1}, we have introduced a U(1) phase factor $\mathrm{e}^{-\mathrm{i}\varphi(r)}$ related to the magnetic field via the vector potential $\boldsymbol{A}$ 
%
%
\begin{equation}\label{flux}
\varphi(r)=\frac{2\pi}{\phi_0}\int_{\alpha}^{r} d\boldsymbol{r}\cdot\boldsymbol{A}(\boldsymbol{r})\; ,
\end{equation}
%
%
where $\phi_0=h/e$ is the flux quantum. This phase eventually leads to the Aharonov-Bohm (AB) effect~\cite{Aharonov_1959,Batelaan_2009} in closed loops. The expression in Eq.~\eqref{flux} is proportional to the circulation of the vector
potential between vertex $\alpha$ and point $r$.~\cite{Kottos:1999,Vidal:2000} The second phase factor in Eq.~\eqref{wavefunction1} is a SU(2) phase due to the RSOC. In closed loops it leads to the Aharonov-Casher (AC) effect~\cite{Casher_1984} (electromagnetic dual of the AB effect) arising from the spin precession driven by the Rashba field.~\cite{Bercioux:2004,Bercioux:2005A,Bercioux:2015}

The probability current corresponding to Hamiltonian~\eqref{HamiltonianABC} is
%
%
\begin{align}\label{current}
j=&-\mathrm{i}\frac{\hbar}{2m^*}\Big\{ \Psi^{\dagger}\frac{\partial\Psi}{\partial r}-\frac{\partial\Psi^{\dagger}}{\partial r}\Psi+2\mathrm{i}\frac{e}{\hbar}\left(\hat{\boldsymbol{\gamma}}\cdot\boldsymbol{A}\right)\Psi^{\dagger}\Psi \\ \nonumber
&+2\mathrm{i}k_{\mathrm{R}}\Psi^{\dagger}\left[(\hat{\boldsymbol{\gamma}}\times\hat{\boldsymbol{z}})\cdot\boldsymbol{\sigma}\right]\Psi\Big\}.
\end{align}
%
%
This expression hints that the probability current is not conserved by the continuity of the derivative of the wavefunction. However, the continuity of the \emph{extended} derivative
%
%
\begin{equation}\label{covariantderivative}
\frac{\partial}{\partial r}\rightarrow D=\frac{\partial}{\partial r}+\mathrm{i}\frac{e}{\hbar}\hat{\boldsymbol{\gamma}}\cdot\boldsymbol{A}+\mathrm{i}k_{\mathrm{R}}(\hat{\boldsymbol{\gamma}}\times\hat{\boldsymbol{z}})\cdot\boldsymbol{\sigma}.
\end{equation}
%
%
does ensure the conservation of the probability current.

Once the wavefunction of all QWs is written as in Eq.~\eqref{wavefunction1}, the conservation of probability current using the extended derivative is imposed at the vertices. Solving the resulting system of equations provides the value of the spinors at all the vertices $\boldsymbol{\Psi}_{\alpha}$, and $\boldsymbol{\Psi}(r)$ by extension.

\subsection{Case of RSOC \& Zeeman field}\label{Rashba+Zeeman}

We now consider a QW subject to RSOC and to an in-plane Zeeman field pointing in the direction $\hat{\boldsymbol{\alpha}}$, where $\boldsymbol{B}=B(\cos\alpha,\sin\alpha,0)$. The system Hamiltonian reads
%
%
\begin{equation}\label{Hamiltonian}
\hat{\mathcal{H}}=\frac{\boldsymbol{p}^{2}}{2m^*}+\frac{\hbar k_{\mathrm{R}}}{m^*}(\boldsymbol{p}\times\boldsymbol{z})\cdot\boldsymbol{\sigma}+\mu\boldsymbol{B}\cdot\boldsymbol{\sigma}\; ,
\end{equation}
%
%
with $\mu$ the Bohr magneton. Unlike the RSOC term, the Zeeman term does not depend on momentum and thus it breaks time-reversal symmetry. 

Using the spinorial wavefunction ansatz of Eq.~\eqref{spinor}, the Hamiltonian can be cast into the following matrix:
%
%
\begin{equation}\label{Hamiltoniank}
\hat{\mathcal{H}}=
  \left(\begin{array}{cc}
    \frac{\hbar^2k^2}{2m^*} & \mathcal{M}^{*} \\
    \mathcal{M} & \frac{\hbar^2 k^2}{2m^*}
  \end{array}\right),
\end{equation}
%
%
with $\mathcal{M}=(\mu B\cos{\alpha}+\frac{\hbar^2 k_{\mathrm{R}}k}{m}\sin{\gamma})+\mathrm{i}(\mu B\sin{\alpha}-\frac{\hbar^2 k_{\mathrm{R}}k}{m}\cos{\gamma})$. The matrix \eqref{Hamiltoniank} has the following eigenvalues and eigenvectors
%
%
\begin{equation}\label{energyzeeman}
\epsilon_{\pm}=\frac{\hbar^2k^2}{2m^*}\pm|\mathcal{M}| \,,
\end{equation}
%
%
%
%
\begin{equation}\label{eigenfunctionzeeman}
|\boldsymbol{v}_{\pm}\rangle = \frac{1}{\sqrt{2}}
  \left(\begin{array}{cc}
    \mathrm{e}^{-\mathrm{i}\theta/2} \\
    \pm\mathrm{e}^{\mathrm{i}\theta/2}
  \end{array}\right)\; ,
\end{equation}
%
%
with $\theta$ the argument of $\mathcal{M}$. The eigenvectors $|\boldsymbol{v}_{\pm}\rangle$ lie within the $xy$-plane and remain constant along the wire.

%
%
\begin{figure}[!h]
    \centering
    \includegraphics[width=\columnwidth]{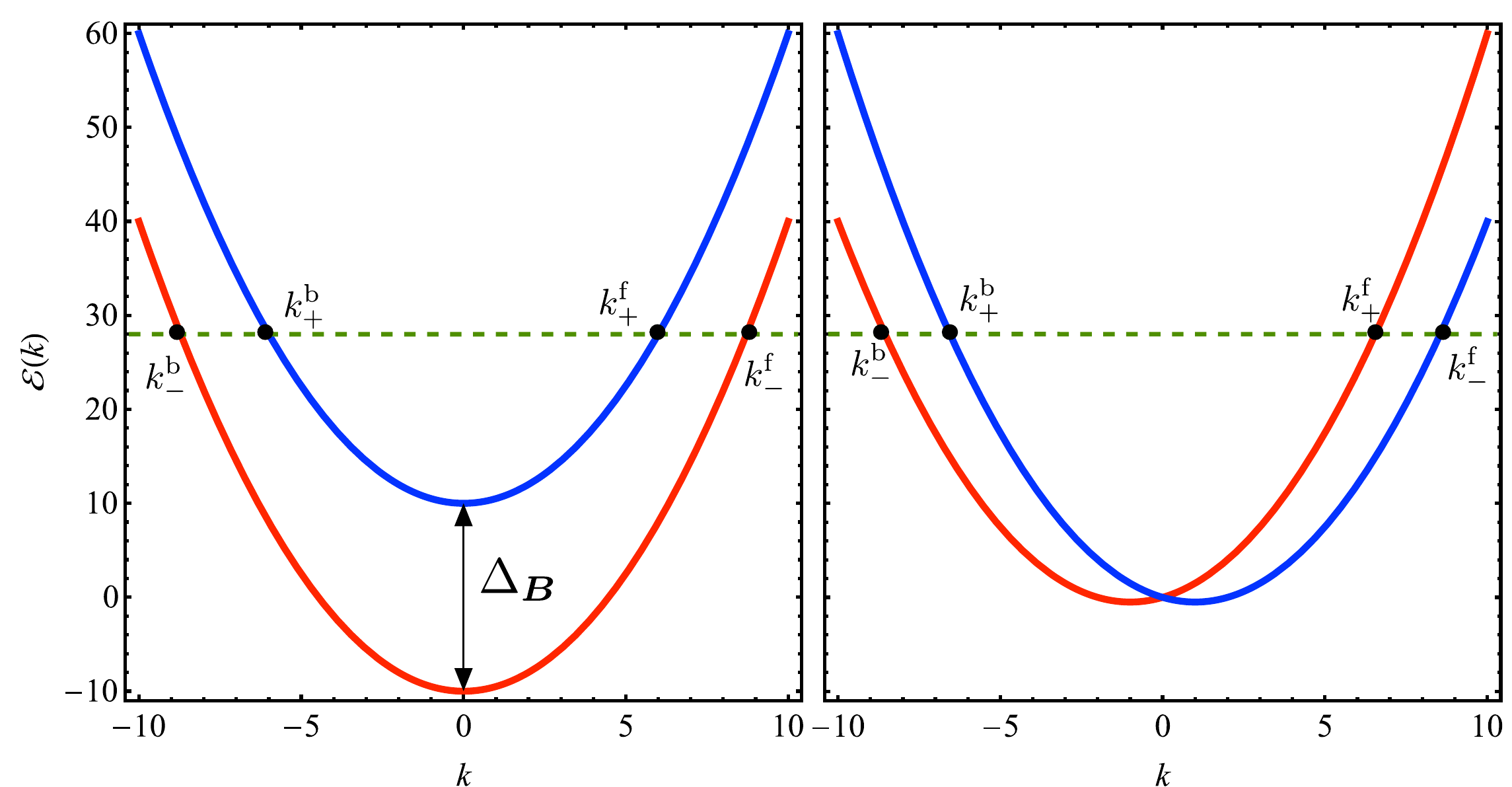}
\caption{\label{energybands} Energy spectrum as a function of momentum for $k_\mathrm{SO}=0$ (left) and $B=0$ (right). In the left panel, $\Delta_{\boldsymbol{B}}$ is the pseudo-gap opened by the Zeeman field at zero momentum. The four propagating states at fixed energy are labelled by $k_\pm^\mathrm{f/b}$.}
\end{figure}
%
%
The energy spectrum of the quantum wire is given by Eq.~\eqref{energyzeeman}, and consists of two energy bands due to the spin splitting induced by the RSOC and Zeeman interactions. The energy bands for the two limiting cases ($k_{\mathrm{R}}=0$ and $B=0$) are shown in Fig.~\ref{energybands}. It is common knowledge that the dispersion relation in the presence of a Zeeman field is shifted vertically for opposite spins, while the RSOC introduces a horizontal shift.~\cite{Bercioux:2015} In general, the interplay between both fields results in a more complex spectrum.

The method described in this section is valid when the Fermi energy $E_{\mathrm{F}}$ lies above the pseudogap $\Delta_{\boldsymbol{B}}=2\mu |\boldsymbol{B}|$, namely, the splitting induced by the Zeeman field at $\bm{k}=0$. In this situation, there are four available propagating states for a given energy, two forward and two backward. We label the momenta and arguments corresponding to the four states as $k_{\pm}^\mathrm{f/b}$ and $\theta_{\pm}^\mathrm{f/b}$, respectively. The superscript f/b indicates forward/backward propagation and the subscript $\pm$ indicates the energy band. A state with a defined propagation direction can be written as
%
%
\begin{equation}\label{gestate}
\boldsymbol{\Psi}(r)=\boldsymbol{\hat{\mathcal{R}}}\mathrm{e}^{\mathrm{i}\bar{k} r}\boldsymbol{\Psi}(0)\,. 
\end{equation}
%
%
Here we have introduced the spin evolution matrix $\boldsymbol{\hat{\mathcal{R}}}$ relating the value of the wavefunction at point $r$ to its initial value at $r=0$. We find an analytic formula for the spin evolution matrix, given by [see Appendix~\ref{appendix_R} for its derivation]
%
%
\begin{align}\label{R}
\boldsymbol{\hat{\mathcal{R}}}=\frac{1}{\cos\left(\frac{\Delta\theta}{2}\right)}\begin{pmatrix}
  \cos{\frac{\Delta k r-\Delta \theta}{2}} & \mathrm{i}\mathrm{e}^{\mathrm{-i}\bar{\theta}}\sin{\frac{\Delta k r}{2}} \\
  \mathrm{i}\mathrm{e}^{\mathrm{i}\bar{\theta}}\sin{\frac{\Delta k r}{2}} & \cos{\frac{\Delta k r+\Delta \theta}{2}}
  \end{pmatrix}.
\end{align}
%
%
For the sake of simplicity, we have suppressed here the superscripts f/b. Additionally, we have introduced the elements $\bar{k}=\frac{k_{+}+k_{-}}{2}$, $\Delta k=k_{+}-k_{-}$, $\bar{\theta}=\frac{\theta_{+}+\theta_{-}}{2}$ and $\Delta \theta=\theta_{+}-\theta_{-}$. Note that in general these four quantities will be different for states propagating forward or backward. Equation~\eqref{R} is one of the original results of this work since it allows for the exact treatment of hybrid RSOC and Zeeman fields within the quantum network method.

In general, the wavefunction of a QW will be a linear combination of the  available counter-propagating waves,
%
%
\begin{subequations}\label{finalstate}
\begin{align}
\boldsymbol{\Psi}(r)=&\boldsymbol{\Psi}^\mathrm{f}(r)+\boldsymbol{\Psi}^\mathrm{b}(r)\label{finalstatea} \\ 
=&\boldsymbol{\hat{\mathcal{R}}}^\mathrm{f}\mathrm{e}^{\mathrm{i}\bar{k}^\mathrm{f} r}\boldsymbol{\Psi}^\mathrm{f}(0)+\boldsymbol{\hat{\mathcal{R}}}^\mathrm{b}\mathrm{e}^{\mathrm{i}\bar{k}^\mathrm{b} r}\boldsymbol{\Psi}^\mathrm{b}(0)\;. \label{finalstateb}
\end{align}
\end{subequations}
%
%
The spinors $\boldsymbol{\Psi}^\mathrm{f}(0)$ and $\boldsymbol{\Psi}^\mathrm{b}(0)$ are unknown constants that are fixed by applying the  boundary conditions.

When the  Zeeman field is zero, the forward and backward elements satisfy the relations $\bar{k}^{\mathrm{b}}=-\bar{k}^{\mathrm{f}}$, $\Delta k^{\mathrm{b}}=-\Delta k^{\mathrm{f}}$, $\bar{\theta}^{\mathrm{b}}=\bar{\theta}^{\mathrm{f}}+\pi$ and $\Delta\theta^{\mathrm{f}}=\Delta\theta^{\mathrm{b}}=0$.
After rearranging $\boldsymbol{\Psi}^{\mathrm{f/b}}(0)$ in terms of $\boldsymbol{\Psi}(0)$ and $\boldsymbol{\Psi}(\ell)$, the wavefunction can be written as
%
%
\begin{align}\label{genwf}
\boldsymbol{\Psi}(r)=\frac{\boldsymbol{\hat{\mathcal{R}}}(r)}{\sin(k\ell)}&[\sin{k(\ell-r)}\boldsymbol{\Psi}(0)\nonumber \\
&+\sin{(kr)}\boldsymbol{\hat{\mathcal{R}}}^{-1}(\ell)\boldsymbol{\Psi}(\ell)]\;,
\end{align}
%
%
where now the spin evolution matrix $\boldsymbol{\hat{\mathcal{R}}}$ coincides with the SU(2) phase factor in Eq.~\eqref{wavefunction1}. 
When the RSOC is zero and we have only the Zeeman term, which breaks time-reversal symmetry, we can still express the wavefunction with a structure similar to Eq.~\eqref{finalstateb}, with the spin rotation matrix that now reads:
%
%
\begin{equation}
    \boldsymbol{\hat{\mathcal{R}}}(r)= \mathrm{e}^{\mathrm{i} \hat{\mathbf{\alpha}}\cdot\boldsymbol{\sigma} \frac{\Delta k}{2}r }\,.
\end{equation}
%
%

We note that a crossing of the energy band  occurs when the Zeeman field is perpendicular to the wire and its modulus reaches to the critical value $B=\frac{\hbar^2 k k_{\mathrm{R}}}{m \mu}$. Under this condition, the effective magnetic field created by the RSOC cancels with the in-plane magnetic field, so the only contribution to the energy comes from the kinetic term. This situation is similar to spin-helix effect arising in systems with Rashba and Dresselhaus SOCs of equal strength.~\cite{Schliemann:2003A,Schliemann:2003B,Koralek_2009} The effective magnetic field due to the spin-fields vanishes for a certain momentum, so the Hamiltonian becomes spin-independent and an energy band crossing occurs. In this case, $\mathcal{M}=0$, so the angles $\theta_{\pm}$ for the two-fold degenerate solutions are not defined. A careful analysis shows that the spin evolution matrix is equal to the identity matrix. This comes as no surprise, since the SU(2) terms in the Hamiltonian cancel each other out. Therefore, the spatial evolution of the state is simply given by the dynamic phase factor $\mathrm{e}^{\mathrm{i}k r}$ that arises from the kinetic term.

Equations \eqref{R} and \eqref{finalstate} are the key step to generalize the quantum network method when an in-plane Zeeman field is applied. Boundary conditions are then applied at the vertices of the wire in order to obtain $\boldsymbol{\Psi}^{\mathrm{f/b}}(0)$, and $\boldsymbol{\Psi}(r)$ by extension. The Zeeman term does not contribute any additional term to the extended derivative, so it is given by~\eqref{covariantderivative} with $\boldsymbol{A}=0$.

\subsection{Formalism for quantum transport}\label{transportformalism}

To study the transport properties of  quantum networks, we attach semi-infinite input and output leads to the vertices of the network.~\cite{Vidal:2000,Bercioux:2004,Bercioux:2005B} Each lead consists of a single-mode QW with two spin channels.~\footnote{This approach has been generalized to include multiple modes by introducing mode mixing at the nodes, see Refs.~[\onlinecite{Khee-Kyun-Voo:2006,Andersen_2018}].} The leads are not subjected to any interaction, so they are characterized at zero temperature by a wave vector $k$ and a Fermi energy $E_\text{F}=\hbar^2k^2/2m$. We assume that the leads are connected to uncorrelated reservoirs, so that there are no phase relationships among electrons in different channels.~\cite{Imry_1997}

In a system with $N_{\mathrm{in}}$ ($N_{\mathrm{out}}$) input (output) channels, if an electron is injected through input channel $\sigma$ with wavenumber $k$, the wavefunction alongside the channels can be written as
%
%
\begin{subequations}
\begin{align}
\Psi_{\text{in},\sigma'}(r)&=\mathrm{e}^{\mathrm{i}k r}\delta_{\sigma'\sigma}+r_{\sigma'\sigma}\mathrm{e}^{-\mathrm{i}k r},\label{input}\\
\Psi_{\text{out},\sigma'}(r)&=t_{\sigma'\sigma}\mathrm{e}^{\mathrm{i}k r},\label{output}
\end{align}
\end{subequations}
%
%
where $r$ is the position measured from the edge (negative for input leads and positive for output leads). Here, $r_{\sigma'\sigma}$ and $t_{\sigma'\sigma}$ are the channel-resolved reflection and transmission amplitudes, respectively. 
The indices $\sigma$ and $\sigma'$ specify both the lead and the spin state of the channel. 
We define the total transmission and reflection coefficients of a channel $\sigma'$ as~\cite{Scheid_2007,Bercioux:2015}
%
%
\begin{subequations}
\begin{align}
    T_{\sigma'}&=\sum_{\sigma}\left|t_{\sigma'\sigma}\right|^2\,, \\
    R_{\sigma'}&=\sum_{\sigma}\left|r_{\sigma'\sigma}\right|^2\, ,
\end{align}
\end{subequations}
%
%
where the sum runs over the input channels. The total transmission (reflection) is given by the sum of the transmission (reflection) coefficients of the output (input) channels,
%
%
\begin{subequations}
\begin{align}
        T=&\sum_{\sigma'}T_{\sigma'}=\sum_{\sigma\sigma'}\left|t_{\sigma'\sigma}\right|^2,\\
        R=&\sum_{\sigma'}R_{\sigma'}=\sum_{\sigma\sigma'}\left|r_{\sigma'\sigma}\right|^2.
\end{align}
\end{subequations}
%
%
The probability conservation (unitarity of the scattering matrix) imposes that $T+R=1$.

The zero-temperature conductance $G$ based on the Landauer formula reads:~\cite{datta_1997}
%
%
\begin{equation}\label{landauer_conductance}
    G=\frac{e^2}{h}\mathrm{Tr}\: \left[tt^{\dagger}\right]=\frac{e^2}{h} T.
\end{equation}
%
%
It is clear from the previous expression that the conductance is bounded by the number input channels, such that $G \leq N_{\mathrm{in}} e^2/h$.

The derivatives of the wavefunction in the leads must also be taken into account when imposing the conservation of probability current. In an isolated quantum network, by imposing the continuity of the wavefunction and the conservation of the probability current we obtain a set of linear homogeneous equations where the variables are the values of the wavefunction at the vertices. This allows us to study the spectral properties of the quantum network via a secular equation.~\cite{Kottos:1999,Gnutzmann:2006}

When adding the external leads, the energy of the system is fixed by the Fermi energy of the leads $E_\text{F}$. The transmission and reflection coefficients can be written in terms of the values of the wavefunction at the contacts. Due to the first term in the right-hand side of Eq.~\eqref{input}, the set of equations becomes inhomogeneous, with a unique solution for $T$ and $R$.

If there is no Zeeman field, the wavefunction of the network is described by the values it takes at the vertices $\boldsymbol{\Psi}_\alpha$ [see Eqs.~\eqref{wavefunction1}]. For each input (output) lead there are two reflection (transmission) coefficients, one per spin channel. To satisfy the single-valuedness of the wavefunction at the vertices connected to external leads, one can write the reflection and transmission coefficients of the leads as a function of $\boldsymbol{\Psi}_\alpha$. The number of variables of the problem is then equal to the number of vertices $V$. At each vertex the sum of the outgoing extended derivatives must be equal to zero, so there are $V$ equations that impose the continuity of probability current. These equations fix the values of $\boldsymbol{\Psi}_\alpha$, and consequently the reflection/transmission coefficients.

Importantly, when the Zeeman field is finite, the wavefunction is described by spinors $\boldsymbol{\Psi}_{\alpha\beta}^\mathrm{f}(0)$ and $\boldsymbol{\Psi}_{\alpha\beta}^\mathrm{b}(0)$, which specify the wavefunction of a bond at one of its endpoints [see~Eqs.~\eqref{finalstate}]. The subscripts indicate that the QW is connected to vertices $\alpha$ and $\beta$. Together with the reflection/transmission coefficients, there are $2N+N_{\mathrm{ext}}$ unknown variables, where $N$ is the number of edges of the quantum network and $N_{\mathrm{ext}}$ is the number of input/output leads. For a vertex $\alpha$ connected to $N_\alpha$ edges, we can write $N_\alpha-1$ equations that impose the single-valuedness of the wavefunction. The total number of equations that verify the continuity of the wavefunction at the edges of quantum network are $2N+N_{\mathrm{ext}}-V$. In addition, at each vertex the sum of the outgoing extended derivatives must be equal to zero. In total there are $2N+N_{\mathrm{ext}}$ equations that fix the values of the spinors and the transmission/reflection coefficients.

For a generic vertex $\alpha$, the continuity of the probability current reads
%
%
\begin{equation}\label{probability_current}
\sum_{\langle\alpha,\beta\rangle} \left. D\boldsymbol{\Psi}_{\alpha,\beta}(r) \right|_{r=0}=0\; ,
\end{equation}
%
%
where the sum $\sum_{\langle\alpha,\beta\rangle}$ runs over all vertices $\beta$ which are connected to $\alpha$.

Equation~\eqref{probability_current} can be expressed in terms of $\boldsymbol{\Psi}_{\alpha\beta}^\mathrm{f}(0)$ and $\boldsymbol{\Psi}_{\alpha\beta}^\mathrm{b}(0)$ using Eq.~\eqref{finalstateb}. In this case, the equation for the internal vertices is
%
%
\begin{equation}
\sum_{\delta\in\{\mathrm{f,b}\}}\sum_{\langle\alpha,\beta\rangle} \boldsymbol{M}_{\alpha,\beta}^{\delta}\boldsymbol{\Psi}_{\alpha,\beta}^{\mathrm{\delta}}(0)=0\; .
\end{equation}
%
%
In the case where the QW is subject to RSOC and a Zeeman field, 
%
%
\begin{align}
\boldsymbol{M}_{\alpha,\beta}^{\delta}=&\mathrm{i}\bar{k}^{\delta}+\frac{\Delta k^{\delta}}{2}\tan\frac{\Delta\theta^{\delta}}{2}\sigma_z\nonumber\\
+&\mathrm{i}\left(\frac{\Delta k^{\delta}}{2}\frac{\hat{\boldsymbol{\theta}}^{\delta}}{\cos\frac{\Delta\theta^{\delta}}{2}}+k_{\mathrm{R}}(\hat{\boldsymbol{\gamma}}\times\hat{\boldsymbol{z}})\right)\cdot\boldsymbol{\sigma}\; ,
\end{align}
%
%
where $\hat{\boldsymbol{\theta}}^{\delta}=(\cos{\theta^\delta},\sin{\theta^\delta},0)$.

Consider a quantum network with a single input and output leads. If an electron with spin $\sigma$ is injected along the input lead, the equations for the external vertices read
%
%
\begin{subequations}\label{current_continuity2}
\begin{align}
\sum_{\substack{\delta\in\{\mathrm{f,b}\} \\ \langle\alpha,\beta\rangle}} \boldsymbol{M}_{\alpha,\beta}^{\delta}\boldsymbol{\Psi}_{\alpha,\beta}^{\mathrm{\delta}}(0)=&\mathrm{i}k\boldsymbol{\chi}_{\sigma}-\mathrm{i}k\sum_{\sigma'}r_{\sigma'\sigma}\boldsymbol{\chi}_{\sigma'},\label{current_continuity2a}\\
\sum_{\substack{\delta\in\{\mathrm{f,b}\} \\ \langle\alpha,\beta\rangle}} \boldsymbol{M}_{\alpha,\beta}^{\delta}\boldsymbol{\Psi}_{\alpha,\beta}^{\mathrm{\delta}}(0)=&-\mathrm{i}k\sum_{\sigma'}t_{\sigma'\sigma}\boldsymbol{\chi}_{\sigma'}\label{current_continuity2b}.
\end{align}
\end{subequations}
%
%

%
%
\begin{figure*}
    \centering
    \includegraphics[width=\textwidth]{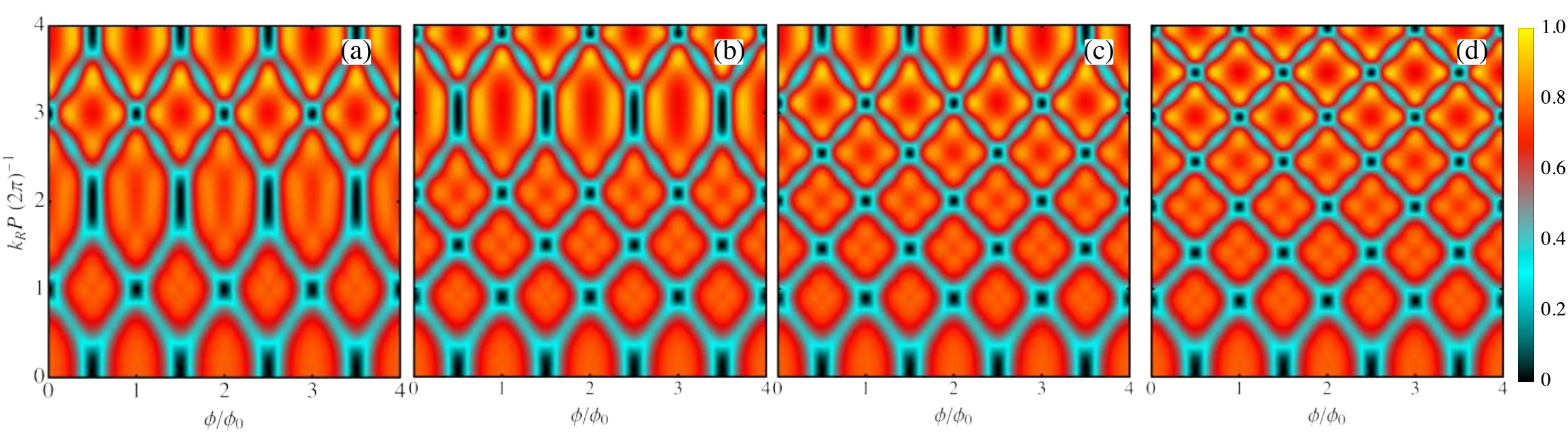}
    \caption{\label{conductance1} Average conductance $\langle G\rangle_k$ in units of $2e^2/h$ for various polygons: (a) square, (b) hexagon, (c) octagon and (d) ring, subject to a magnetic flux $\phi/\phi_0$ and RSOC $k_\mathrm{R} P/(2\pi)$.}
\end{figure*}
%
%
\noindent The coefficients $r_{\sigma'\sigma}$ and $t_{\sigma'\sigma}$ can be expressed as a linear combination of $\boldsymbol{\Psi}_{\alpha\beta}^\mathrm{\delta}(0)$ by applying the continuity of the wavefunction. Together with the equations that impose the continuity of the wavefunction of the internal edges, we obtain an inhomogeneous  system  of linear equations with $4N$ variables (2 per spinor): the inhomogeneous term arises due to the first term on the rhs of Eq.~\eqref{current_continuity2a}. The system can be  solved numerically to obtain the value of the spinor of each bond at the local coordinate $r=0$. Once $\boldsymbol{\Psi}_{\alpha\beta}^\mathrm{\delta}(0)$ are obtained, it is straightforward to compute $r_{\sigma'\sigma}$ and $t_{\sigma'\sigma}$. Furthermore, Eq.~\eqref{finalstateb} provides the value of the wavefunction at any given point. In the case where the Zeeman term is zero, the boundary conditions for the leads reduce to the known cases in Ref.~[\onlinecite{Vidal:2000,Bercioux:2004,Bercioux:2005A}]:
%
%
\begin{equation}
\boldsymbol{M}_{\alpha\alpha}\boldsymbol{\Psi}_{\alpha}+\sum_{\langle\alpha,\beta\rangle}\boldsymbol{M}_{\alpha\beta}\boldsymbol{\Psi}_{\beta}=0\; .
\end{equation}
%
%

\section{Results}\label{results}

In this section we study the transport properties of different polygons using the formalism we have introduced in the previous section. We consider a series of regular polygons of constant perimeter $P$ with an even number of vertices. Each polygon is connected to an input and an output field-free leads at opposite vertices | see  Fig.~\ref{polygons}(a)-\ref{polygons}(d). 
We evaluate the system conductance from the transmission probability applying the Landauer-B\"{u}ttiker formalism~\cite{Buttiker_1985} as dictated by Eq.~\eqref{landauer_conductance}. Taking the number of edges to infinity, the series of regular polygons converges to a circle, so we recover the conductance for a ring. For the case in which only RSOC is present, it was shown in Ref.~[\onlinecite{Bercioux:2005B}] that this numerical procedure coincides with the analytical results for rings.~\cite{Frustaglia:2004}

In order to match the conditions present in mesoscopic transport experiments, our numerical model must satisfy a series of constraints. The first one is the so-called semiclassical limit requiring the electronic wavelength to be much smaller than the system's size. This condition can be written in terms of the electronic wavenumber and the polygon's perimeter as        
%
%
\begin{equation}
k \gg \frac{2\pi}{P}\; .
\end{equation}
%
%
This limit justifies the interpretation of the system's conductance in terms of carrier interference along classical-like propagating paths.~\footnote{For reviews on semiclassical theory, see e.g., Refs.~[\onlinecite{Richter:2000,Jalabert_2000}]} Moreover, it has been observed that slow carriers with smaller wavenumbers are more prone to decoherence.~\cite{Nagasawa:2013} 

The wavenumber is also constrained by the size of the polygon's edges. On the one hand, a polygonal geometry results noticed by the carriers on the condition that their wavelength is much smaller than the edges length, so that multiple wavelengths fit in one edge. Otherwise, the electron would not ``see" the polygon's vertices. On the other hand, in the specific case of ring modeling, the requirement is just the opposite: the wavelength must be much larger than the edges, instead, so that each vertex is hardly noticed by the carriers and the polygon can be interpreted as a ring:
%
%
\begin{equation}
k \ll \frac{2\pi N}{P}=\frac{2\pi}{L}\; ,
\end{equation}
%
%
where $N$ is the number of edges of the polygon used to simulate the ring, and $L$ is the length of each edge. Clearly, the larger the number of edges, the fittest the ring's polygonal model. Moreover, for similar reasons, the spin precession length  $L_{\mathrm{SO}}$ should be much longer than edges as well.~\footnote{For a recent discussion in semiclassical terms, see the Supplemental Material in Ref.~[\onlinecite{Wang_2019}].}
%
%
\begin{figure*}
    \centering
    \includegraphics[width=\textwidth]{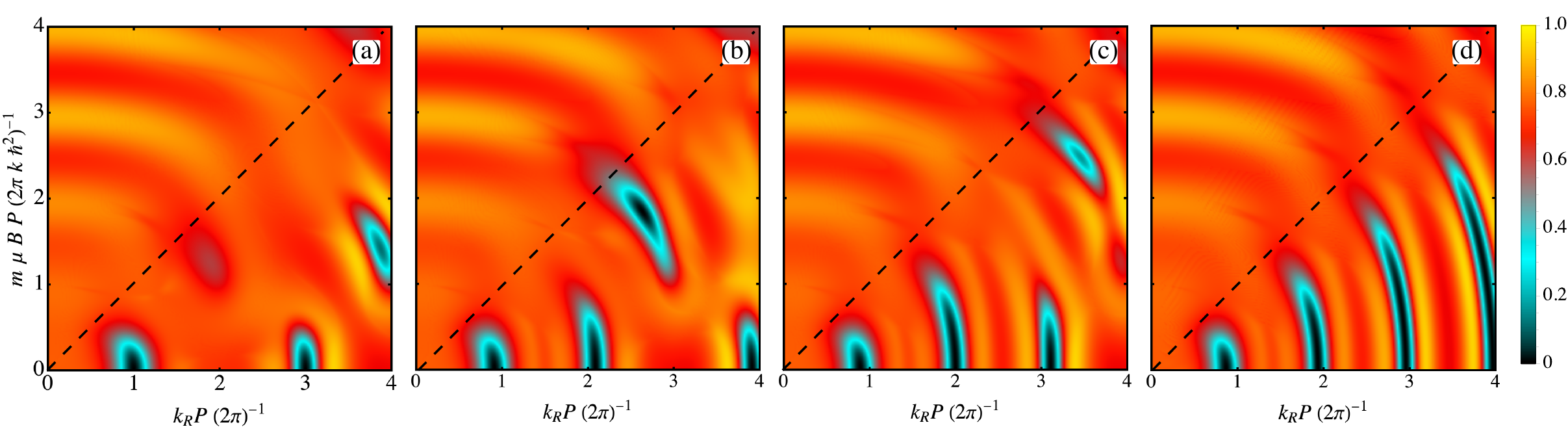}
    \caption{\label{conductance}Average conductance $\langle G\rangle_k$ in units of $2e^2/h$ for $\alpha=0$ for various polygons: (a) square, (b) hexagon, (c) octagon and (d) ring. The dashed line corresponds to the critical line, where $B_{\mathrm{SO}}=B$.}
\end{figure*}
%
%
\subsection{Case of RSOC \& magnetic flux}

Let us begin by discussing the transport properties of polygons subject to magnetic flux and RSOC. 
We present in Fig.~\ref{conductance1} the average conductance $\langle G \rangle_k$ for different polygons. The average, performed over a small $k$-window around the Fermi wavenumber $k_{\rm F}$ of incoming carriers, smooths out the energy-dependent oscillations of the conductance due to Fabry-P\'erot-like interference.~\cite{Frustaglia:2004,Bercioux:2004} This reproduces the situation found in low-(but-still-finite)-temperature transport experiments. The simulations of ring geometries are carried out by using polygons with 100 edges.

The  results presented in Fig.~\ref{conductance1} are a perfect example for highlighting the difference between an Abelian and a non-Abelian gauge field, due to the orbital magnetic field and the RSOC, respectively. 
The conductance shows periodic AB oscillations, where the period is the flux quantum $\phi_0$ for all the polygons. The maxima correspond to the constructive quantum interference of the electrons travelling through different paths. For example, in the Rashba field free limit, the constructive interference occurs for integer multiples of the flux quantum. The AB phase acquired by an electron when moving around the polygon is $2\pi\phi/\phi_0$. Adding the contributions of all the possible paths gives rise to the interference pattern. The phase acquired by a particle moving through the shortest possible paths (clockwise and counterclockwise paths) will have the same magnitude, but opposite sign. For integer multiples of the flux quantum, the phase difference between the two paths is an integer multiple of $2\pi$ resulting in a constructive interference. However, the conductance is not exactly $2e^2/h$. The longer paths will also contribute to the transmission amplitude, each of which have a different dynamical phase.
On the other side, for half integer multiples of the flux quantum, the contribution of the two opposite paths to the transmission amplitude are in counterphase, so the conductance drops to zero.

The RSOC modifies the phase acquired by the electrons when travelling through the polygon; this leads to a shift of the position of the conductance maxima with respect to the magnetic flux. For instance, at $k_{\mathrm{R}}P=2\pi$ the conductance maxima appear at half integer multiples of $\phi_0$ for all polygons, while the conductance vanishes for integer multiples of the flux quantum. The additional phase arises form the electron spin precession around the effective magnetic field arising from the RSOC, \emph{i.e.} it acquires a non-Abelian SU(2) phase resulting in the Aharonov-Casher effect.~\cite{Casher_1984} The spin-phase gathering is controlled by two different scales: the perimeter and the edge lengths. This is reflected on the oscillations of the conductance, which show broader and narrower maxima for different values of $k_{\mathrm{R}}P$ associated to two different frequencies.~\cite{Bercioux:2005B} The periodicity of the broader maxima is related to the edge lengths, where the period is $N\pi$. This period tends to infinity as the number of edges tends to infinity, so the broad maxima disappear for the ring, apart from the one located at the origin.

The periodicity of the narrow maxima is related to the length of the perimeter, therefore, it has a weaker dependence on the number of edges of the polygon. The quasiperiod ranges from $4\pi$ for the case of the square to $2\pi$ as the number of edges and the RSOC strength increase. Oscillations of period $2\pi$ are identified with the adiabatic limit: when the dimensionless RSOC tends to infinity, we obtain the adiabatic limit in which the spin is aligned with the effective magnetic field during transport, and Berry phases arise~\cite{Bercioux:2005B,Frustaglia:2004}. Adiabatic spin transport is never really achieved in polygons, where vertices act as spin-scattering centers due to the abrupt change of direction of the RSOC at the vertices of polygons. The results of Figs.~\ref{conductance1}(a) and ~\ref{conductance1}(d) show a qualitative agreement with the experimental observations in ring~\cite{Koening_2006,Bergsten_2006,Nagasawa_2012} and square structures.~\cite{Koga_2006} Still, a full agreement of the magnetoconductance periodicities require the introduction of disorder since the experiments present a periodicity halving due to the Altshuler-Aronov-Spivak (AAS) effect, a manifestation of the AB effect in the presence of dominant time-reversed-paths interference.~\cite{Altshuler_1981} Such a periodicity halving will be reproduced in Sec. \ref{disorder_case} by implementing a numerical model for disorder. 

\subsection{Case of RSOC \& Zeeman field}
%
%
\begin{figure*}[!t]
\centering
  \includegraphics[width=0.95\textwidth]{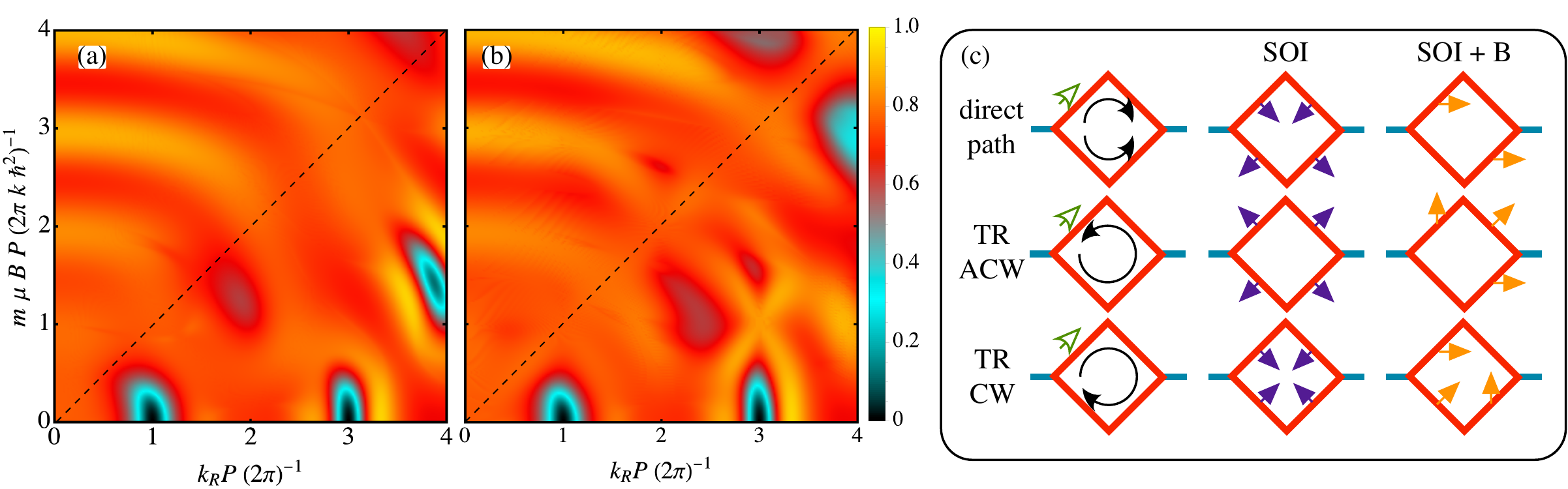}
\caption{\label{alpha0}  Average conductance $\langle G\rangle_k$ of a square for a Zeeman field with an orientation of $\alpha=0$ (a) and $\alpha=\pi/4$ (b). (c) Total magnetic field in a square for direct clockwise and counterclockwise (first row) paths and time-reversed (second \& third rows) paths when $B_{\mathrm{SO}}=B$ and $\alpha=\pi/4$. The green arrows represents the Zeeman field, the purple ones the field associated to the RSOC and the yellow ones the effective field due to the combination of Zeeman and RSOC.}
\end{figure*}
%
%
In this section we study the interplay between Zeeman and AC phases using the method described in Sec.~\ref{Rashba+Zeeman}. The quantum wires of the network are subject to RSOC and an in-plane Zeeman field, which breaks time-reversal symmetry. The averaged conductance for different polygons is shown in Fig.~\ref{conductance}. The dashed line represents the critical line, the points where the applied Zeeman field is equal in magnitude to the effective magnetic field created by the RSOC, $B_{\mathrm{SO}}=\frac{\hbar^2 k k_{\mathrm{R}}}{m \mu}$, which can be obtained from direct comparison between the second and third terms in Eq.~\eqref{Hamiltonian}. The dimensionless RSOC and Zeeman couplings are given by $k_{\mathrm{R}} P/(2\pi)$ and $m\mu B P/(2\pi k \hbar^2)$, respectively.

For rings, Fig.~\ref{conductance}(d), the AC oscillations shift to weaker values of the SOC field as the applied Zeeman field increases.These results are qualitatively consistent with the conductance shift observed experimentally by Nagasawa \textit{et al.} in InGaAs-based quantum ring arrays, attributed to a geometric-phase manipulation.~\cite{Nagasawa:2013} However, in the experiment the periodicity of the magnetoresistance oscillations is halved due to the AAS effect. We fully recover this halving by introducing disorder | see Sec.~\ref{disorder_case}.

The RSOC field $B_{\mathrm{SO}}$ is radial to the ring, while the in-plane Zeeman field is homogeneous. Both fields lie on the $xy$ plane, so the solid angle subtended by the magnetic field in parameter space corresponding to the Berry phase depends on whether the total magnetic field encircles the origin or not.~\cite{Saarikoski:2015} For simplicity, the Zeeman field is applied along the direction of the positive $x$-axis. For $B<B_{\mathrm{SO}}$, the solid angle is $\Omega=2\pi$, corresponding to a Berry phase $\pi$. However, for $B>B_{\mathrm{SO}}$ the solid angle vanishes, so that the Berry phase is $0$.

The topology of the field texture, which coincides with the spin-eigenstate texture in the adiabatic limit, changes when the critical line is traversed. The results in Fig.~\ref{conductance} show that the effects of this transition manifest in the conductance oscillations even in a nonadiabatic regime  --- small $k_\mathrm{R} P/(2\pi)$.

Strongly non-adiabatic spin textures allow for a transition of the topological properties of the spin eigenmodes when an in-plane Zeeman field is applied. These properties are characterized by the winding parity of the spin eigenmodes.~\cite{Wang_2019} As it has been stated, for the ring this topological transition occurs when the applied Zeeman field is equal to the effective spin-orbit field. However, the AC oscillations for polygons with a finite number of edges show a sign reversal for much lower values of the Zeeman field.

Figure~\ref{conductance} shows the conductance for a Zeeman field direction $\alpha=0$. However, the conductance pattern shows a strong dependence of $\alpha$ for intermediate values of $B$. While the ring has a continuous rotational symmetry, polygons only remain invariant under rotations of an integer multiple of $2\pi/N$. In addition, they have $N$ symmetry planes, as well as inversion symmetry. The input and output leads break the rotational symmetry of polygons. Numerical calculations show that the only remaining symmetry planes are the horizontal and vertical planes. Therefore, the conductance will be the same for angles $\alpha$, $\alpha'=-\alpha$ and $\alpha''=\pi-\alpha$. Introducing elastic disorder to the system removes these symmetry planes, although the inversion center remains, so that the conductance is the same for $\alpha$ and $\alpha'''=\alpha+\pi$.

The conductance of a square for two orientations of the Zeeman field is shown in Fig.~\ref{alpha0}. For small values of the Zeeman field strength, the RSOC field contributes the most to the total magnetic field, so the oscillation pattern is almost identical in both cases. In the small RSOC limit, the conductance is also very similar for both orientations: for $k_{\mathrm{R}}=0$, the total magnetic field is equal to the applied Zeeman field, so it is homogeneous along the ring. The only effect on the energy spectrum is the splitting of the energy bands into two parabolas with opposite spin direction (see Fig.~\ref{energybands}). The phase acquired by the electrons following the two shortest possible paths (the direct clockwise and counterclockwise paths) is the same, although the contribution of the longer paths results in the interference effect that gives rise to the oscillatory effect.

%
%
\begin{figure}[!t]
  \centering
  \includegraphics[width=0.95\columnwidth]{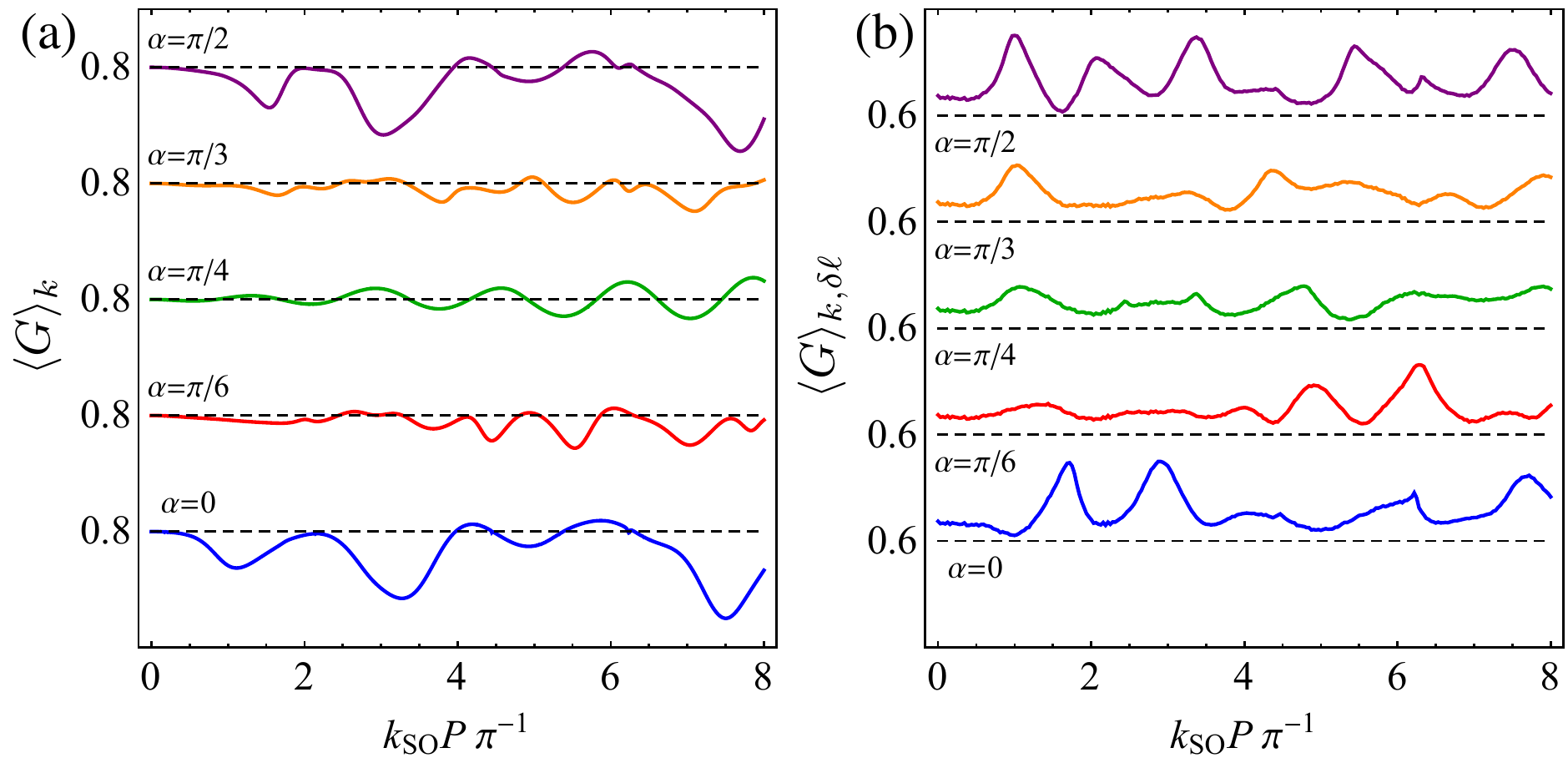}
\caption{\label{alphamultiple} Average conductance $\langle G\rangle_k$ as a function of the direction of the Zeeman field $\alpha$ along the critical line for the  square (a) without disorder and (b) with disorder. The disorder is implemented so as explained in Sec.~\ref{disorder_case}. Disorder decreases the average conductance due to the random value of the dynamical phase.  The conductance traces are vertically shifted of a constant $0.2$ factor for clarity.}
\end{figure}
%
%

The main difference between Fig.~\ref{alpha0}(a) and \ref{alpha0}(b) is found around the critical line, where the oscillations of the conductance for $\alpha=\pi/4$ are smoothed, see Fig.~\ref{alphamultiple}(a). As explained in Sec.~\ref{Rashba+Zeeman}, the energy bands of a wire become degenerate when the Zeeman and RSOC strength are the same and the Zeeman field is perpendicular to one of the edges of the polygon. The RSOC field is perpendicular to the motion of the electron, so in some edges, it cancels with the applied Zeeman field. For $\alpha=\pi/4$, the Zeeman field is perpendicular to two opposite edges of the square, as it can be seen in Fig.~\ref{alpha0}(c). Therefore, electrons with any spin orientation propagating along these edges will not precess, and the interference effect between the upper and lower paths will be mitigated. In addition, the total magnetic field along the remaining two edges is the same, so the phase acquired by electrons following the direct clockwise and counterclockwise paths is the same. Higher-order contributions, due to paths that go several times around the polygons, are responsible for the non-constant conductance along the critical line.

The conductance along the critical line for several values of $\alpha$ is shown in Fig.~\ref{alphamultiple}. For the square, as $\alpha$ approaches the $\alpha=\pi/4$ direction, the oscillations become less pronounced. Moreover, the oscillations show a more regular pattern, with a steadily increasing amplitude.

Similar effects of mitigation of the spin procession due to RSCO are present in polygons with larger number of edges as well. However, the oscillatory pattern for the corresponding values of $\alpha$ do not resemble the regular pattern obtained for the square. For a fixed perimeter $P$, as the number of edges of the polygon increases, the length of the edges become smaller, so the edge where the spin does not process has a smaller contribution towards the interference.

\subsection{Disordered case}\label{disorder_case}
%
%
\begin{figure}[!t]
  \centering
  \includegraphics[width=0.95\columnwidth]{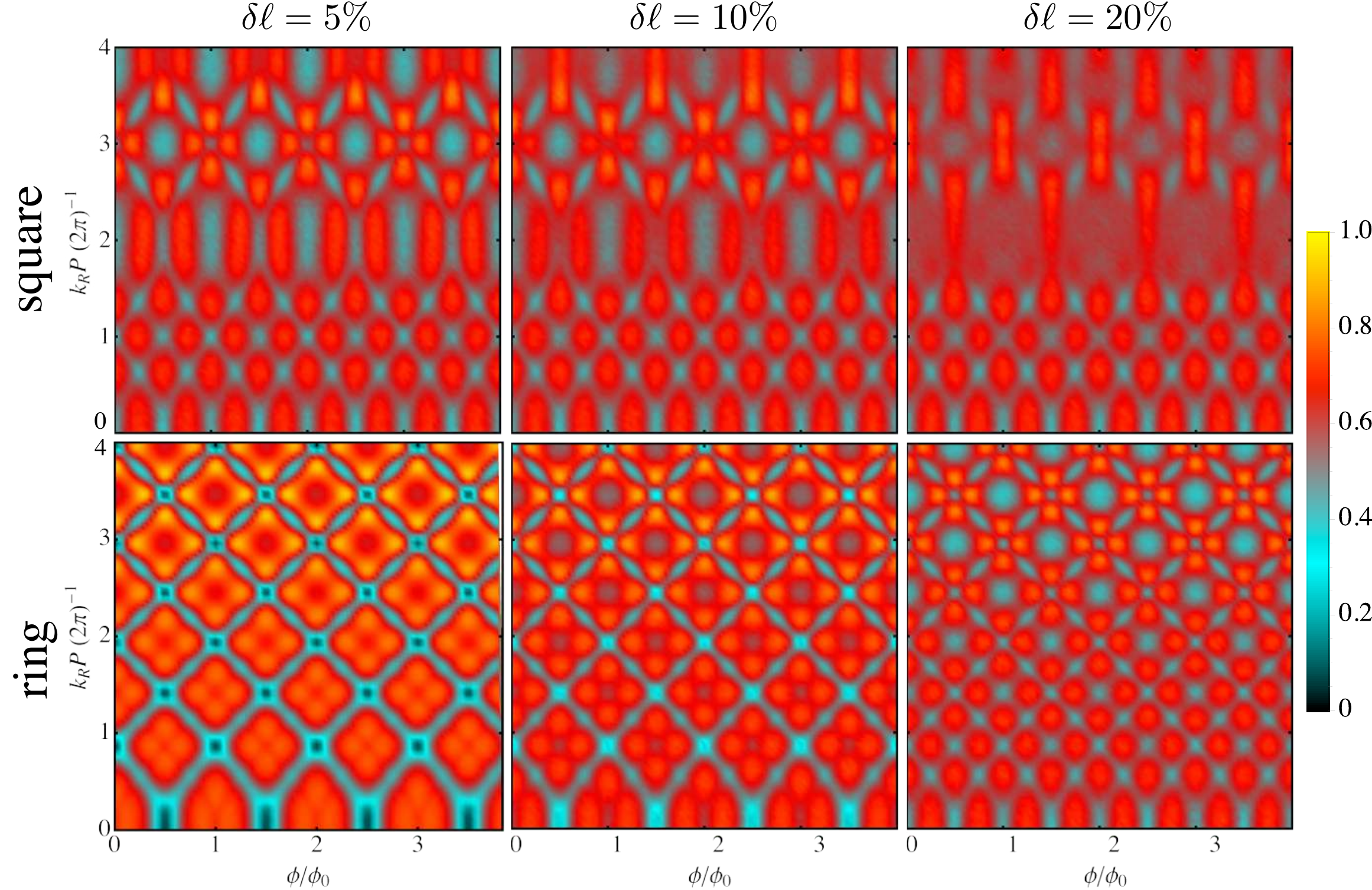}
\caption{\label{disorderABC}Average conductance $\langle G\rangle_{k,\delta\ell}$ as a function of the orbital magnetic field and of the RSOC in the presence of disorder for the cases of the square (upper row) and of the ring (lower row) obtained with the quantum network method. The three columns refer to different strengths of the length fluctuations: $\delta \ell=5\%$, $10 \%$ and $20 \%$.}
\end{figure}
%
%
In this section, we show how disorder affects the results we have presented in the previous sections. In general, disorder is inevitable in nanostructures and we need to account for its effects.
Within the quantum network formalism, disorder can be introduced in several ways, \emph{e.g.} randomly distributed pointlike scatterers, or more generally, the random elastic scattering matrix along the edges.
We implement disorder following the  scheme proposed in Refs.~[\onlinecite{Vidal:2000,Bercioux:2004,Bercioux:2005A,Bercioux:2005B}]: to simulate arbitrary shifts of the wavefunction phase we implement random fluctuation of the length of each edge while preserving the orientation of the edges and keeping the perimeter $P$ of the polygon constant. We introduce therefore a length fluctuation parameter $\delta \ell$. These fluctuations account for the geometric imperfections present in experiments with arrays of rings and squares.\cite{Nagasawa_2012,Nagasawa:2013,Nagasawa:2018,Wang_2019} In the following we will focus on the case $k\delta\ell\sim1$, which can be reached in the semiclassical regime without a sizable modification of the loops areas such that the periodicity with respect to the magnetic flux is preserved.~\cite{Vidal:2000} Moreover, we introduce an additional energy average as done in the disorder-free case to boost numerical convergence.

Contrary to the standard Anderson-like disorder, this model of elastic disorder in a single-loop quantum network will not produce a wavefunction localization, but will lead to a reduction of the amplitude of oscillations of the average conductance. We will show that one of the main effects of disorder is to double the period of the oscillations of the average conductance of the system, and to lift  the complete destructive interference.~\cite{Bercioux:2005A,Vidal:2000,Bercioux:2005B} The doubling of the period of oscillations corresponds to the so-called AAS oscillations~\cite{Altshuler_1981} when considering the effect of the magnetic field or AAS-like oscillations when considering the RSOC.~\cite{Bergsten_2006,Nagasawa_2012} The emergence of the AAS oscillations in quantum networks with this type of disorder has previously been demonstrated by analysing the Fourier amplitude of the conductance versus the magnetic flux in Ref.~[\onlinecite{Vidal:2000}].
%
%
\begin{figure}[!t]
  \centering
  \includegraphics[width=0.95\columnwidth]{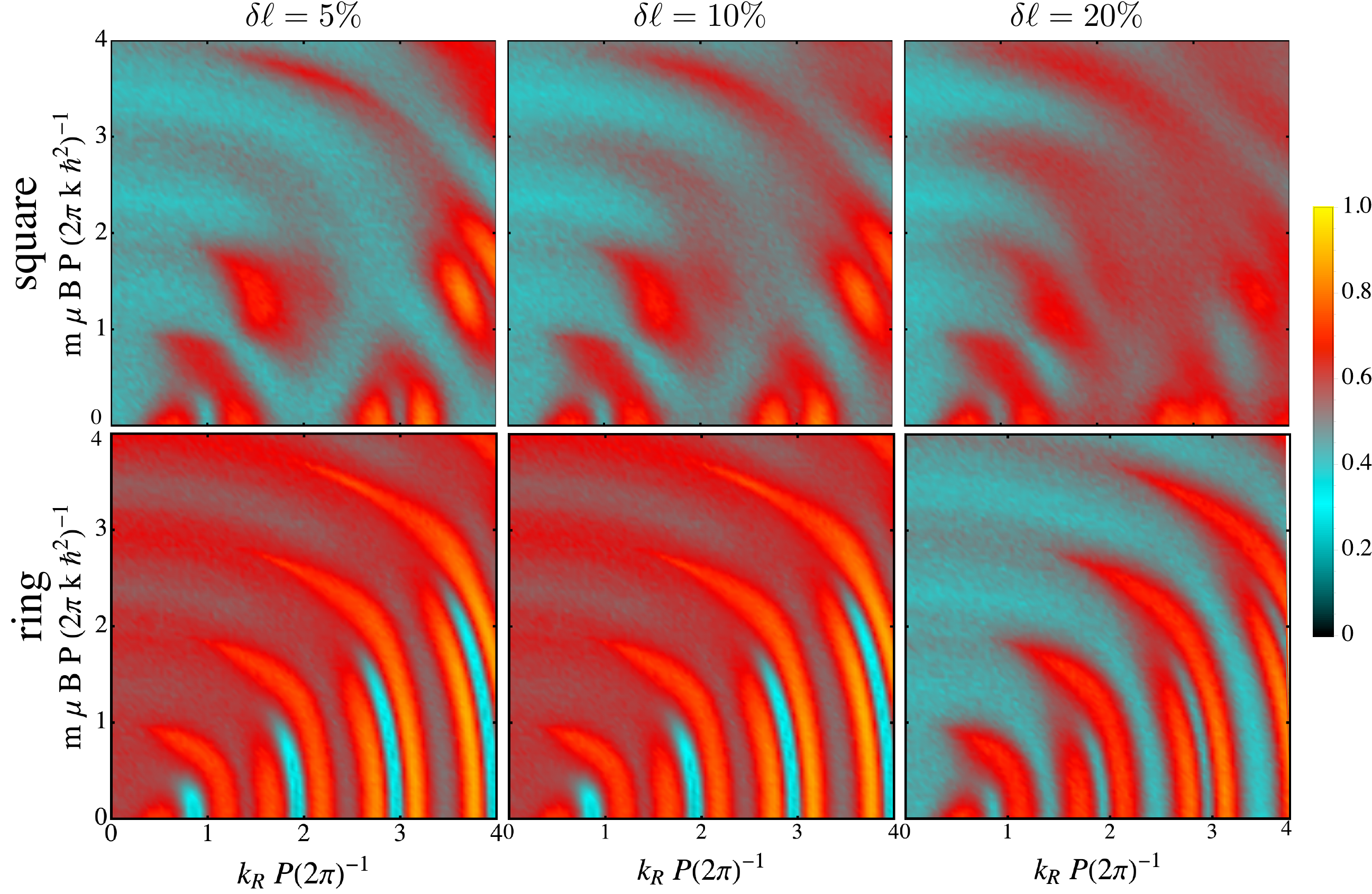}
\caption{\label{disorderACZ}Average conductance $\langle G\rangle_{k,\delta\ell}$ as a function of the in-plane Zeeman term and of the RSOC in the presence of disorder for the cases of the square (upper row) and of the ring (lower row) obtained with the quantum network method. The three columns refer to different strengths of the length fluctuations: $\delta \ell=5\%$, $10 \%$ and $20 \%$.}
\end{figure}
%
%
These results were in complete agreement with the experimental finding presented in Ref.~[\onlinecite{Naud_2001}]. Similar results were also presented in the presence of RSOC in one- and two-dimensional disordered quantum networks.~\cite{Bercioux:2004,Bercioux:2005A}
These quantum oscillations are dominated by the time-reversed paths.~\cite{Nagasawa:2013,Wang_2019} For the sake of simplicity, in the following we present results for the case of disorder only considering the polygonal structures that are experimentally more relevant: the square and the ring --- see Fig.~\ref{disorderABC} for the case of AB and AC interference, and Fig.~\ref{disorderACZ} for the case of AC and Zeeman field. The results shown in Figs.~\ref{disorderABC} and \ref{disorderACZ} | especially those for $\delta \ell = 20\%$, to be compared to Figs. \ref{conductance1} and \ref{conductance}, respectively | present a frequency doubling emerging from semiclassical time-reversed paths interference, which agree with the corresponding experimental findings for rings \cite{Nagasawa_2012,Nagasawa:2013} and squares.~\cite{Koga_2006,Wang_2019} Additionally, by following the development of the interference pattern as the disorder strength increases, we further notice from Figs.~\ref{disorderABC} and \ref{disorderACZ} that the ring geometry appears to be less susceptible to show the frequency doubling (i.e., relatively stronger disorder is required by rings as compared to squares). This is likely due to the role played by the square vertices as spin-carrier scatterers.

\section{Validation of the quantum network method within the tight-binding approach}\label{res_tb}
%
%
\begin{figure*}
  \centering
  \includegraphics[width=0.85\textwidth]{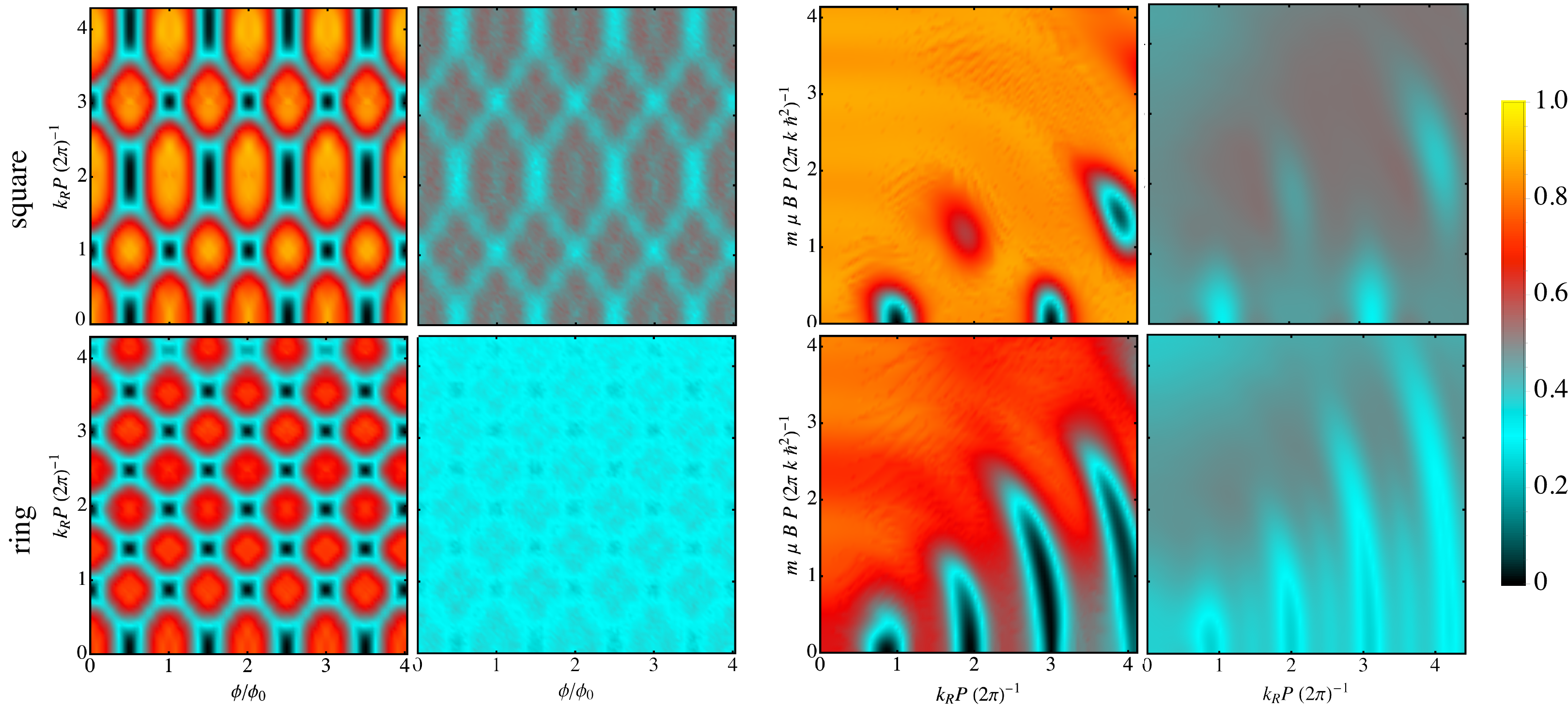}
\caption{\label{fig_TB_results}
Summary of the tight-binding results for the average conductance as a function of the magnetic flux $\phi/\phi_0$ and RSOC $k_R P(2\pi)^{-1}$ (left) and as a function of the RSOC $k_R P(2\pi)^{-1}$ and Zeeman momentum $m \mu B P(2\pi k \hbar^2)^{-1}$ (right). Both cases are given for squares (upper rows) and circles (lower rows), for clean (left columns) and disordered systems (right columns).}
\end{figure*}
%
%
In this section we present results of a fully numerical approach, using a tight-binding model. We use \textsc{Kwant}, a \textsc{python} package facilitating tight-binding transport simulations.~\cite{Groth_2014} The typical system we run simulations on are depicted in Figs.~\ref{polygons}(e) and~\ref{polygons}(f). We choose the units such that \(\hbar = m^* = e = 1\). 
The Hamiltonian for a square lattice with RSOC~\cite{Mireles_2001} is
%
%
\begin{eqnarray}
H_\mathrm{RSOC} = \frac{\mathrm{i} \lambda}{2a} \sum_i \left( c_{i}^\dagger \sigma_x c_{i+\hat y } - c_{i}^\dagger \sigma_y c_{i+\hat x } \right) + \mathrm{H.c.}
\end{eqnarray}
%
%
where \(\lambda\) is the RSOC strength, \(a\) is the lattice size, \( c_{i}^\dagger / c_{i}\) are the creation/annihilation operators, and \(\sigma_i , \, i \in \{x,y,z\}\) are the Pauli matrices.
The orbital magnetic field is implemented via a Peierls substitution $t \longrightarrow t \mathrm{e}^{\mathrm{i} \varphi}$. 
Following  the notation of the analytical section, the Hamiltonian for the Zeeman field is written as
%
%
\begin{eqnarray}
H_\mathrm{Zeeman} = \frac{\mu |\boldsymbol{ B}|}{2 a^2} \sum_i  c_{i}^\dagger c_{i} (\sigma_x \cos \alpha  + \sigma_y \sin \alpha ).
\end{eqnarray}
%
%
We stick to narrow systems here, in order for the system to contain only a small number of energy bands (few modes). For calculating the average conductance we integrate over  an energy window containing several level spacings, provided single mode occupancy is observed. The results can be seen in Fig.~\ref{fig_TB_results}, where the left block gives the results for the case including an orbital magnetic field and the right block is for the case including a Zeeman field. We find good qualitative agreement with the analytical results. Deviations from the analytical results can come from multiple sources. In the simulations we have to make a compromise between taking a thin wire, which can host only few modes, and a thicker wire, through which the electrons can move more easily without scattering. Results presented here have been run for an approximate width $W=3a$ of the system (leads and edges), giving $P/W \approx 60$. In wider systems the amplitude of the conductance oscillations is reduced, probably due to coupling to higher modes.

For simulations with disorder we use Anderson type disorder, meaning each lattice site $i$ is subjected to an onsite energy $\varepsilon$ which is randomly chosen from an interval $[-U_0/2, \, U_0/2]$. The Anderson Hamiltonian can be written as
%
%
\begin{eqnarray}
H_\mathrm{Anderson} = \frac{U_0}{2 a^2} \sum_i \varepsilon_i c_{i}^\dagger c_{i} 
\end{eqnarray}
%
%
where $a$ is the lattice constant. In our comparison between disordered systems, one has to take into account that Anderson disorder also results in localization of the wavefunction. One can therefore not expect a direct correspondence between the analytical and numerical results  with respect to the magnitude of the conductance. As can be observed in the right panels of each block in Fig.~\ref{fig_TB_results}, the conductance decreases substantially, because the disorder strength we used is strong, namely $U_0 = 0.8 t$, which is $20$\% of $4t$. Here we have averaged over 100 disorder configurations. Despite the overall lowering of the conductance, we do observe the same pattern as in the case of the disordered quantum networks, which means that the general behavior as a function of the external parameters is the same. Moreover, the numerical results correctly reproduce the results obtained in experiments.~\cite{Koga_2006,Nagasawa_2012,Nagasawa:2013,Wang_2019} For weaker disorder strengths, the conductance sticks to higher values, but the period doubling is not clearly observed.

\section{Conclusions and Outlook}\label{conclusions}
In this work, we have presented a generalization of the quantum network method to model one-dimensional networks subject to hybrid field textures produced by the combined action of out-of-plane and in-plane magnetic fields and Rashba spin-orbit couplings. Contrary to previous studies,~\cite{Nagasawa:2013} this method does not rely on perturbation theory, so it is valid for systems with high Zeeman fields. In addition, while other methods rely on spin-related phases accumulated by the carriers between input and output leads by following geometric paths with a finite number of windings, this method calculates the exact wavefunction of the electron along the edges. More specifically, we have studied the interplay of the non-Abelian phases introduced by the Aharonov-Casher effect and the in-plane Zeeman field in one-dimensional polygons, in addition to the customary Abelian phase produced by the flux associated to a perpendicular magnetic field. We have considered polygonal structures with an increasing number of vertices in order to \emph{simulate} a ring structure by sending to infinity the number of vertices. We have shown that in the case of the Rashba spin-orbit coupling, there is a double dependence of the conductance both on the perimeter of the system and the number of vertices itself. In the case of the interplay between the two non-Abelian phases due to the in-plane Zeeman field and the Rashba spin-orbit coupling, our results for the conductance are in agreement with the experimental observations.~\cite{Nagasawa:2013,Wang_2019} Furthermore, we have supported our quantum network approach comparing our results with the ones obtained by a standard tight-binding approach. 
The quantum network method can also be used to obtain the spectrum and the eigenfunctions of the isolated polygons via the corresponding secular equation,~\cite{Kottos:1999,Gnutzmann:2006} which can be used to calculate the geometric and dynamical phases. It is also an attractive technique to study complex structures which can become computationally very demanding within the tight-binding approach. The quantum network method has been successfully employed for investigating the chaotic properties of complex networks, results obtained are in full agreement with random matrix theory.~\cite{Gnutzmann:2006}  The theory presented here offers an extension to this line of research. Furthermore, the quantum network method can be additionally extended to study spin transport on the surface state of three dimensional topological insulators so to describe the Aharonov-Casher physics observed in Bi$_2$Se$_3$ square-ring interferometers.~\cite{Qu_2011}

\section{Acknowledgements}

We acknowledge useful discussion with Ainhoa Iñiguez. This work was supported by the Spanish Ministerio de Ciencia, Innovaci\'{o}n y Universidades through Projects Nos. FIS2017-82804-P (A.H., T.B. and D.B.) and FIS2017-86478-P (D.F.), and by the Transnational Common Laboratory $Quantum-ChemPhys$ (D.B.).

\appendix

\section{Derivation of the spin evolution matrix}\label{appendix_R}

In this Appendix we derive the spin evolution matrix defined in Eq.~\eqref{R}, which relates the value of the wavefunction at point $r$ to its initial value at $r=0$. For this purpose, we expand the spin up/down states $|\mathrel{\mspace{-7mu}} \uparrow \mathrel{\mspace{-8mu}} / \mathrel{\mspace{-8mu}} \downarrow\rangle$ into the base formed by the available states {$\{\lvert\boldsymbol{v}_\pm\rangle\}$} [see Eq.~\eqref{eigenfunctionzeeman}]. These states are not orthogonal, but satisfy the relation
%
%
\begin{equation}\label{orthogonality}
\langle\boldsymbol{v}_\alpha | \boldsymbol{v}_\beta\rangle=\frac{1}{2}\left(\mathrm{e}^{\mathrm{i}(\theta_\alpha-\theta_\beta)/2}+\alpha\beta\mathrm{e}^{-\mathrm{i}(\theta_\alpha-\theta_\beta)/2} \right)\; .
\end{equation}
%
%

In order to expand a generic vector into a non orthogonal basis $\lvert\boldsymbol{u}\rangle=c_+\lvert\boldsymbol{v}_+\rangle+c_-\lvert\boldsymbol{v}_-\rangle$, we make use of the Gram matrix:
%
%
\begin{equation}
\boldsymbol{M}=
  \begin{pmatrix}
  \langle \boldsymbol{v}_+|\boldsymbol{v}_+\rangle &  \langle \boldsymbol{v}_+|\boldsymbol{v}_-\rangle \\
  \langle \boldsymbol{v}_-|\boldsymbol{v}_+\rangle &  \langle \boldsymbol{v}_-|\boldsymbol{v}_-\rangle
  \end{pmatrix}\; .
\end{equation}
%
%
Coefficients $c_\pm$ satisfy the relation
%
%
\begin{equation}
\boldsymbol{u} = \boldsymbol{M} \boldsymbol{c} \; ,
\end{equation}
%
%
where
%
%
\begin{equation}
 \boldsymbol{u} = \binom{\langle \boldsymbol{v}_+|\boldsymbol{u}\rangle}{\langle \boldsymbol{v}_-|\boldsymbol{u}\rangle},\quad \boldsymbol{c} = \binom{c_+}{c_-}\; .
\end{equation}
%
%

Taking the inverse of the Gram matrix, we obtain the coefficients $c_\pm$. Spin up/down states can therefore be written as
%
%
\begin{subequations}\label{updown}
\begin{equation}
\lvert\uparrow\rangle=\frac{1}{\sqrt{2}\cos{\frac{\theta_+-\theta_-}{2}}}\left(\mathrm{e}^{\mathrm{i}\theta_-/2}\lvert\boldsymbol{v}_+\rangle+\mathrm{e}^{\mathrm{i}\theta_+/2}\lvert\boldsymbol{v}_-\rangle\right)\; ,
\end{equation}
\begin{equation}
\lvert\downarrow\rangle=\frac{1}{\sqrt{2}\cos{\frac{\theta_+-\theta_-}{2}}}\left(\mathrm{e}^{-\mathrm{i}\theta_-/2}\lvert\boldsymbol{v}_+\rangle-\mathrm{e}^{-\mathrm{i}\theta_+/2}\lvert\boldsymbol{v}_-\rangle\right)\; .
\end{equation}
\end{subequations}
%
%

Knowing that eigenvectors~\eqref{eigenfunctionzeeman} acquire a dynamical phase $\mathrm{e}^{\mathrm{i}kr}$ when travelling a distance $r$, the propagation of the states $|\mathrel{\mspace{-7mu}} \uparrow \mathrel{\mspace{-8mu}} / \mathrel{\mspace{-8mu}} \downarrow\rangle$ is given by
%
%
\begin{subequations}
\begin{equation}
\lvert f\rangle_{\uparrow}=\frac{\mathrm{e}^{\mathrm{i}\theta_-/2}\mathrm{e}^{\mathrm{i}k_+r}\lvert\boldsymbol{v}_+\rangle+\mathrm{e}^{\mathrm{i}\theta_+/2}\mathrm{e}^{\mathrm{i}k_-r}\lvert\boldsymbol{v}_-\rangle}{\sqrt{2}\cos{\frac{\theta_+-\theta_-}{2}}}\; ,
\end{equation}
\begin{equation}
\lvert f\rangle_{\downarrow}=\frac{\mathrm{e}^{-\mathrm{i}\theta_-/2}\mathrm{e}^{\mathrm{i}k_+r}\lvert\boldsymbol{v}_+\rangle-\mathrm{e}^{-\mathrm{i}\theta_+/2}\mathrm{e}^{\mathrm{i}k_-r}\lvert\boldsymbol{v}_-\rangle}{\sqrt{2}\cos{\frac{\theta_+-\theta_-}{2}}}\; .
\end{equation}
\end{subequations}
%
%

The states $\lvert f\rangle_{\uparrow/\downarrow}$ can be expanded into their up and down components through the transfer matrix
%
%
\begin{equation}
\boldsymbol{\hat{\mathcal{T}}}=
  \begin{pmatrix}
  \langle \uparrow| f\rangle_{\uparrow} & \langle \downarrow| f\rangle_{\uparrow} \\
  \langle \uparrow| f\rangle_{\downarrow} & \langle \downarrow| f\rangle_{\downarrow}
  \end{pmatrix}\; ,
\end{equation}
%
%
where
%
%
\begin{equation}\label{transfer}
\boldsymbol{\hat{\mathcal{T}}}=\frac{1}{\cos{\frac{\Delta \theta}{2}}}
  \begin{pmatrix}
  \cos{\frac{\Delta kr-\Delta \theta}{2}} & \mathrm{i}\mathrm{e}^{\mathrm{i}\theta}\sin{\frac{\Delta kr}{2}} \\
  \mathrm{i}\mathrm{e}^{\mathrm{-i}\theta}\sin{\frac{\Delta kr}{2}} & \cos{\frac{\Delta kr+\Delta \theta}{2}}
  \end{pmatrix}\mathrm{e}^{\mathrm{i}kr}\; .
\end{equation}
%
%
Here we have introduced the parameters $k=\frac{k_+ + k_-}{2}$, $\Delta k=k_+-k_-$, $\theta=\frac{\theta_+ + \theta_-}{2}$ and $\Delta \theta=\theta_+-\theta_-$. Note that in general these four quantities will be different for states propagating forward or backward.

Let a wavefunction with a defined propagation direction at $r=0$ be given by,
%
%
\begin{equation}
\boldsymbol{\Psi}(0)=\chi_\uparrow\lvert\uparrow\rangle+\chi_\downarrow\lvert \downarrow\rangle\; .
\end{equation}
%
%
The spatial evolution of the state is then given by
%
%
\begin{equation}\label{finalstate2}
\boldsymbol{\Psi}(r)=\chi_\uparrow\lvert f\rangle_\uparrow +\chi_\downarrow\lvert f\rangle_\downarrow\; .
\end{equation}
%
%

Applying Eq.~\eqref{transfer} to Eq.~\eqref{finalstate2}, we write the value of the wavefunction as a spin-rotation of $\boldsymbol{\Psi}(0)$:
%
%
\begin{equation}
\begin{split}
\boldsymbol{\Psi}(r)=&
  \boldsymbol{\hat{\mathcal{R}}}\mathrm{e}^{\mathrm{i}k r}\boldsymbol{\Psi}(0)\\
  =&\frac{1}{\cos{\frac{\Delta \theta}{2}}}
  \begin{pmatrix}
  \cos{\frac{\Delta k r-\Delta \theta}{2}} & \mathrm{i}\mathrm{e}^{\mathrm{-i}\theta}\sin{\frac{\Delta k r}{2}} \\
  \mathrm{i}\mathrm{e}^{\mathrm{i}\theta}\sin{\frac{\Delta k r}{2}} & \cos{\frac{\Delta k r+\Delta \theta}{2}}
  \end{pmatrix}\mathrm{e}^{\mathrm{i}k r}\boldsymbol{\Psi}(0)\; ,
\end{split}
\end{equation}
%
%
where $\boldsymbol{\hat{\mathcal{R}}}$ is the spin evolution matrix, which is equal to the transpose of the transfer matrix excluding the $\mathrm{e}^{\mathrm{i}k r}$ factor.

\bibliography{biblio}

\end{document}